\documentclass[%
reprint,
superscriptaddress,
 amsmath,amssymb,
 aps,
 prl,
]{revtex4-2}

\usepackage{graphicx}% Include figure files
\usepackage{dcolumn}% Align table columns on decimal point
\usepackage{bm}% bold math
\begin{document}

%\preprint{APS/123-QED}

\title{Fermiology and Origin of $\textit{T}_{c}$ Enhancement in a Kagome Superconductor $\textrm{Cs(V}_{1-x}\textrm{Nb}_{x})_{3}\textrm{Sb}_{5}$}% Force line breaks with \\

\author{Takemi Kato}
\thanks{These authors contributed equally to this work.}
\affiliation{Department of Physics, Graduate School of Science, Tohoku University, Sendai 980-8578, Japan}

\author{Yongkai Li}
\thanks{These authors contributed equally to this work.}
\affiliation{Centre for Quantum Physics, Key Laboratory of Advanced Optoelectronic Quantum Architecture and Measurement (MOE), School of Physics, Beijing Institute of Technology, Beijing 100081, P. R. China}
\affiliation{Beijing Key Lab of Nanophotonics and Ultrafine Optoelectronic Systems, Beijing Institute of Technology, Beijing 100081, P. R. China}
\affiliation{Material Science Center, Yangtze Delta Region Academy of Beijing Institute of Technology, Jiaxing, 314011, P. R. China}

\author{Kosuke Nakayama}
\thanks{Corresponding authors:\\
k.nakayama@arpes.phys.tohoku.ac.jp\\
zhiweiwang@bit.edu.cn\\
t-sato@arpes.phys.tohoku.ac.jp}
\affiliation{Department of Physics, Graduate School of Science, Tohoku University, Sendai 980-8578, Japan}
\affiliation{Precursory Research for Embryonic Science and Technology (PRESTO), Japan Science and Technology Agency (JST), Tokyo, 102-0076, Japan}

\author{Zhiwei Wang}
\thanks{Corresponding authors:\\
k.nakayama@arpes.phys.tohoku.ac.jp\\
zhiweiwang@bit.edu.cn\\
t-sato@arpes.phys.tohoku.ac.jp}
\affiliation{Centre for Quantum Physics, Key Laboratory of Advanced Optoelectronic Quantum Architecture and Measurement (MOE), School of Physics, Beijing Institute of Technology, Beijing 100081, P. R. China}
\affiliation{Beijing Key Lab of Nanophotonics and Ultrafine Optoelectronic Systems, Beijing Institute of Technology, Beijing 100081, P. R. China}
\affiliation{Material Science Center, Yangtze Delta Region Academy of Beijing Institute of Technology, Jiaxing, 314011, P. R. China}

\author{Seigo Souma}
\affiliation{Center for Spintronics Research Network, Tohoku University, Sendai 980-8577, Japan}
\affiliation{Advanced Institute for Materials Research (WPI-AIMR), Tohoku University, Sendai 980-8577, Japan}

\author{Fumihiko Matsui}
\affiliation{UVSOR Synchrotron Facility, Institute for Molecular Science, Okazaki 444-8585, Japan}

\author{Miho Kitamura}
\affiliation{Institute of Materials Structure Science, High Energy Accelerator Research Organization (KEK), Tsukuba, Ibaraki 305-0801, Japan}

\author{Koji Horiba}
\affiliation{Institute of Materials Structure Science, High Energy Accelerator Research Organization (KEK), Tsukuba, Ibaraki 305-0801, Japan}
\affiliation{National Institutes for Quantum Science and Technology (QST), Sendai 980-8579, Japan}

\author{Hiroshi Kumigashira}
\affiliation{Institute of Multidisciplinary Research for Advanced Materials (IMRAM), Tohoku University, Sendai 980-8577, Japan}

\author{Takashi Takahashi}
\affiliation{Department of Physics, Graduate School of Science, Tohoku University, Sendai 980-8578, Japan}
\affiliation{Center for Spintronics Research Network, Tohoku University, Sendai 980-8577, Japan}
\affiliation{Advanced Institute for Materials Research (WPI-AIMR), Tohoku University, Sendai 980-8577, Japan}

\author{Yugui Yao}
\affiliation{Centre for Quantum Physics, Key Laboratory of Advanced Optoelectronic Quantum Architecture and Measurement (MOE), School of Physics, Beijing Institute of Technology, Beijing 100081, P. R. China}
\affiliation{Beijing Key Lab of Nanophotonics and Ultrafine Optoelectronic Systems, Beijing Institute of Technology, Beijing 100081, P. R. China}

\author{Takafumi Sato}
\thanks{Corresponding authors:\\
k.nakayama@arpes.phys.tohoku.ac.jp\\
zhiweiwang@bit.edu.cn\\
t-sato@arpes.phys.tohoku.ac.jp}
\affiliation{Department of Physics, Graduate School of Science, Tohoku University, Sendai 980-8578, Japan}
\affiliation{Center for Spintronics Research Network, Tohoku University, Sendai 980-8577, Japan}
\affiliation{Advanced Institute for Materials Research (WPI-AIMR), Tohoku University, Sendai 980-8577, Japan}
\affiliation{International Center for Synchrotron Radiation Innovation Smart (SRIS), Tohoku University, Sendai 980-8577, Japan}

%\date{\today}

\begin{abstract}
Kagome metals $\textrm{AV}_{3}\textrm{Sb}_{5}$ (A = K, Rb, and Cs) exhibit a characteristic superconducting ground state coexisting with charge-density wave (CDW), whereas the mechanisms of the superconductivity and CDW have yet to be clarified.
Here we report a systematic ARPES study of $\textrm{Cs(V}_{1-x}\textrm{Nb}_{x})_{3}\textrm{Sb}_{5}$ as a function of Nb content $x$, where isovalent Nb substitution causes an enhancement of superconducting transition temperature ($T_{c}$) and the reduction of CDW temperature ($T_{\textrm{CDW}}$).
We found that the Nb substitution shifts the Sb-derived electron band at the $\Gamma$ point downward and simultaneously moves the V-derived band around the M point upward to lift up the saddle point (SP) away from the Fermi level, leading to the reduction of CDW-gap magnitude and  $T_{\textrm{CDW}}$.
This indicates a primary role of the SP density of states to stabilize CDW.
The present result also suggests that the enhancement of superconductivity by Nb substitution is caused by the cooperation between the expansion of the Sb-derived electron pocket and the recovery of the V-derived density of states at the Fermi level.

\end{abstract}

\maketitle

The kagome lattice offers a fertile ground to explore exotic quantum phenomena originating from electron correlation and non-trivial band topology.
Band structure of a simple kagome lattice is composed of a flat band over the entire momentum ($\textbf{k}$) region in the hexagonal Brillouin zone (BZ), a Dirac cone at the BZ corner, and a saddle point (SP) at the zone boundary.
Owing to such a unique band structure, the kagome lattice shows various interesting physical properties depending on the position of the Fermi level ($E_{\textrm{F}}$).
When $E_{\textrm{F}}$ is located at around the flat band, ferromagnetism \cite{MielkeJPA1991_1,MielkeJPA1991_2,MielkeJPA1992,TasakiPRL1992} and fractional quantum Hall effect appear \cite{TangPRL2011,SunPRL2011,NeupertPRL2011,WangPRL2011}, whereas topological insulator \cite{GuoPRB2009,WenPRB2010} and magnetic Weyl semimetal \cite{YangNJP2017} phases are realized when $E_{\textrm{F}}$ is situated at around the Dirac point \cite{LiuScience2019,MoraliScience2019,LiuNP2018,KurodaNM2017,NayakSciAdv2016,YinNP2019,LinPRL2018}.
Despite many theoretical predictions for exotic quantum states such as superconductivity and density-wave ordering \cite{YuPRB2012,ShengPRB2013,KieselPRB2012,KieselPRL2013}, materials with $E_{\textrm{F}}$ at/near the SP are rare.

Recently, a family of kagome metals $\textrm{AV}_{3}\textrm{Sb}_{5}$ (AVS; A = K, Rb, Cs) with a V kagome network [Fig. 1(a)] has emerged \cite{OrtizPRM2019} as a new platform to study the physics associated with the SP due to its proximity to $E_{\textrm{F}}$, as revealed by density-functional-theory calculations \cite{OrtizPRM2019,JiangNM2021,TanPRL2021,WangPRB2022,OrtizPRL2020,FuPRL2021,ZhaoNature2021,FuPRB2021,LiPRX2021,ZhouPRB2021} and angle-resolved photoemission spectroscopy (ARPES) \cite{LouPRL2022,NakayamaPRB2021,LiuPRX2021,ChoPRL2021,LuoNCOM2022,KangNP2022,WangarXiv2021}.
AVS commonly exhibits superconductivity ($T_{c}$ = 0.9-2.5 K) and charge-density wave (CDW; $T_{\textrm{CDW}}$ = 78-103 K) \cite{OrtizPRM2021,OrtizPRL2020,YinCPL2021}.
The mechanism has been intensively discussed in terms of the characteristic band structure with the SP at the M point of BZ [Fig. 1(b)]\cite{JiangNM2021,TanPRL2021,WangPRB2022,FuPRL2021,ZhaoNature2021,FuPRB2021,LiPRX2021,ZhouPRB2021,LouPRL2022,NakayamaPRB2021,LiuPRX2021,ChoPRL2021,LuoNCOM2022,KangNP2022,WangarXiv2021}.
For instance, the scattering connecting different SPs would promote unconventional superconducting pairing \cite{ShengPRB2013,NandkishoreNP2021,WuPRL2021}, whereas the same scattering also contributes to the energy gain to stabilize the chiral CDW with an in-plane $2\times2$ periodicity \cite{ChenNature2021,JiangNM2021,ZhaoNature2021,LiangPRX2021,LiPRX2021,WangPRB2021,ShumiyaPRB2021,OrtizPRX2021}.

A promising strategy to study the interplay among superconductivity, CDW, and electronic states is to modulate the band structure by varying key physical parameters that characterize the electronic phase diagram, such as pressure and carrier concentration \cite{DasNJP2016,SunNCOM2016}.
Such attempts have been made in AVS \cite{DuPRB2021,ChenPRL2021,ZhangPRB2021,DuCPB2021,NakayamaPRX2022,SongPRL2021,OeyPRM2022,LiuarXiv2021,YangarXiv2021,QianPRB2021}.
For example, transport measurements under high pressure have clarified the anticorrelation between $T_{c}$ and  $T_{\textrm{CDW}}$ as well as an unconventional double superconducting dome \cite{DuPRB2021,ChenPRL2021,ZhangPRB2021,DuCPB2021}, whereas the relevant band structure is under debate \cite{LaBollitaPRB2021}.
Chemical substitution in crystal is a useful method to tune chemical pressure and carrier concentrations, but such studies are still limited in AVS \cite{NakayamaPRX2022,LiuarXiv2021,YangarXiv2021,OeyPRM2022}.

In this paper, we report ARPES study of Nb-substituted CVS, $\textrm{Cs(V}_{1-x}\textrm{Nb}_{x})_{3}\textrm{Sb}_{5}$ (CVNS; $x$ = 0, 0.03, and 0.07), in which substitution of V ions with isovalent Nb ions leads to the $T_{c}$ enhancement and simultaneously the $T_{\textrm{CDW}}$ reduction [Fig. 1(c)].
We have revealed that Nb-substitution induces the shift of bands characterized by an expansion of the Sb-derived electron pocket and a shrinkage of the V-derived electron pocket forming the SP, the latter of which is well correlated with the CDW suppression.
We discuss implications of the present results in relation to the mechanism of CDW and superconductivity.

First, we present the electronic structure at the highest Nb concentration ($x = 0.07$) (see Supplemental Material for details on the experimental conditions \cite{refSM}).
Figure 1(d) shows the ARPES-intensity plot at $E_{\textrm{F}}$ as a function of $k_{x}$ and $k_{y}$ at $T$ = 120 K (above $T_{\textrm{CDW}}$) measured with 106-eV photons corresponding to the $k_{z}\sim0$ plane \cite{NakayamaPRB2021}.
Photoelectrons in this wide \textbf{k} window were simultaneously collected by using a momentum microscope \cite{MatsuiJJAP2020}.
One can recognize a circular pocket and a large hexagonal pocket centered at $\Gamma$, together with a small triangular pocket at K.
The circular pocket originates from a parabolic electron band, $\alpha$, as recognized in the ARPES intensity along the $\Gamma$KM cut in Fig. 1(e).
This band is attributed to the $5p_{z}$ band of Sb1 atoms embedded in the kagome-lattice plane [Fig. 1(a)] \cite{NakayamaPRB2021,NakayamaPRX2022,JiangNM2021,WangPRB2022,TanPRL2021,FuPRL2021,ZhaoNature2021}.
The hexagonal pocket originates from linearly dispersive bands $\beta$/$\gamma$, and is attributed mainly to the V $3d_{xz/yz}$ orbital \cite{NakayamaPRB2021,NakayamaPRX2022,JiangNM2021,WangPRB2022}.
The triangular pocket is associated with the $\delta$ and $\varepsilon$ bands, which form a Dirac-cone-like dispersion with the Dirac point at ($E_{\textrm{B}})$ $\sim$0.3 eV.
The $\delta$ band is attributed to the V $3d_{xy/x^{2}-y^{2}}$ orbital \cite{NakayamaPRB2021,NakayamaPRX2022,JiangNM2021,WangPRB2022}, and forms the SP slightly above $E_{\textrm{F}}$.
At the M point, there are two other holelike bands, $\zeta$ and $\eta$ topped at around $E_{\textrm{F}}$ and $E_{\textrm{B}}\sim0.45$ eV, respectively, originating from the V $3d_{xz/yz}$ orbital \cite{NakayamaPRB2021,NakayamaPRX2022,JiangNM2021,WangPRB2022}.

\begin{figure}[htbp]
\includegraphics[width=86mm]{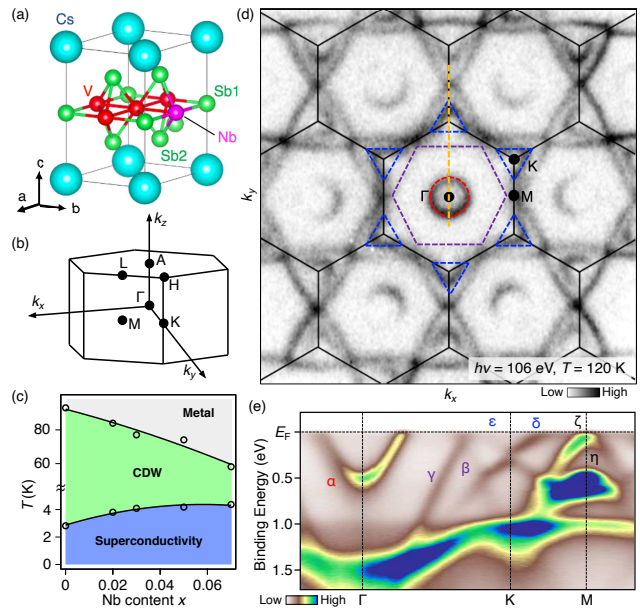}
 \caption{\label{Fig1}
 (a), (b) Crystal structure and bulk Brillouin zone (BZ) of $\textrm{Cs(V}_{1-x}\textrm{Nb}_{x})_{3}\textrm{Sb}_{5}$ (CVNS), respectively.
 (c) Superconducting ($T_{c}$) and CDW ($T_{\textrm{CDW}}$) transition temperatures of CVNS plotted against $x$ \cite{NbCVSZhiwei}.
 (d) ARPES-intensity map at $E_{\textrm{F}}$ plotted as a function of $k_{x}$ and $k_{y}$, measured at $T$ = 120 K with $h\nu$ = 106 eV. Red, purple, and blue dashed lines are guides for the eyes to trace the experimental Fermi surfaces.
 (e) ARPES intensity at $T$ = 120 K measured along a yellow dotted line in (d).}
 \end{figure}

\begin{figure*}[htbp]
\includegraphics[width=165mm]{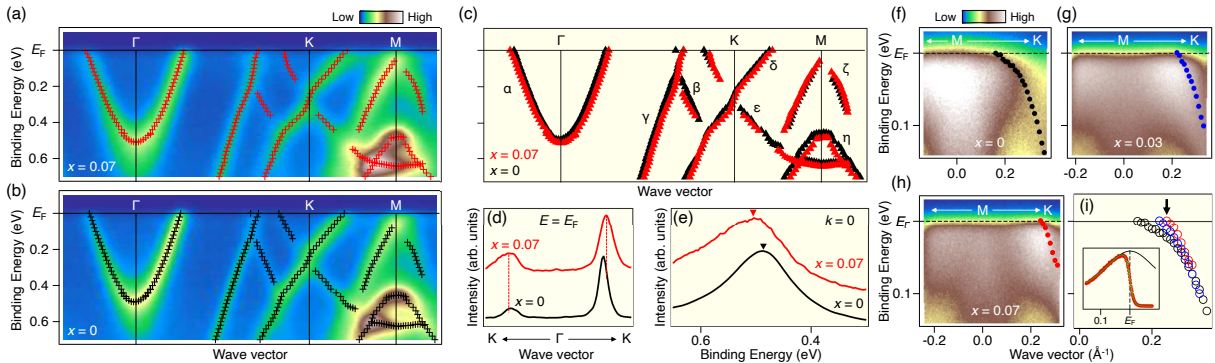}
 \caption{\label{Fig2}
(a) ARPES-intensity plot along the $\Gamma$KM cut for $x = 0.07$ measured at $T$ = 120 K with $h\nu = 106$ eV.
Crosses show experimental band dispersion extracted from the peak position in EDCs and MDCs.
(b) Same as (a) but for $x = 0$.
(c) Comparison of band dispersions between $x = 0.07$ and $x = 0$ [same as red and black crosses in (a) and (b)].
(d), (e) Comparison of MDCs at $E_{\textrm{F}}$ and EDCs at $k_{y} = 0$, respectively, between $x = 0.07$ (red) and $x = 0$ (black).
Red dashed lines in (d) are a guide for the eyes to trace the peak position of $x = 0.07$.
Triangles in (e) indicate the peak position.
(f-h) Plots of ARPES intensity near $E_{\textrm{F}}$ around the M point for $x$ = 0, 0.03, and 0.07, respectively.
Temperature of the measurement was set above $T_{\textrm{CDW}}$ (120, 90, 80 K for $x$ = 0, 0.03, and 0.07, respectively).
(i) Comparison of experimental band dispersions among $x = 0$ (black circles), 0.03 (blue circles), and 0.07 (red circles). Inset shows the representative fitting result (green curve) to the EDC at the $k_{F}$ point for $x = 0.07$ (red circles). The $k_{F}$ point is indicated by black arrow in (i).
The fitting assumes a single Lorentzian peak (black curve) multiplied by the Fermi-Dirac distribution (FD) function convoluted with a resolution function.}
\end{figure*}

Figures 2(a) and 2(b) display a comparison of the ARPES intensity between $x = 0.07$ and $x = 0$ measured along the $\Gamma$KM cut at $T = 120$ K with the energy resolution $\Delta E = 35$ meV.
In both measurements, the electron band $(\alpha)$ at $\Gamma$, the linearly dispersive bands $(\gamma,\varepsilon)$ crossing $E_{\textrm{F}}$, and the $\delta$ band forming a SP at M are commonly resolved.
This indicates that the overall band structure is unchanged after the Nb substitution.
To clarify quantitative differences in the band position, we have determined the experimental band dispersion (red and black crosses), and directly compared them in Fig. 2(c).
At first glance, both band structures for $x$ = 0.07 and 0 well overlap each other.
However, a closer look reveals a finite difference in the energy position of the $\alpha$ band, which shifts downward by $\sim$20 meV at $x = 0.07$ relative to that at $x = 0$.
The shift is also seen by a comparison of MDC at $E_{\textrm{F}}$ in Fig. 2(d) where the peaks for $x = 0.07$ are located outside of the peaks for $x = 0$ (note that the peak width is slightly wider in $x = 0.07$ due to the increase of impurity scattering by Nb substitution).
Moreover, as shown by the EDCs at $\Gamma$ in Fig. 2(e), the bottom of $\alpha$ band moves toward higher $E_{\textrm{B}}$ by $\sim$20 meV with Nb substitution (note that the reproducibility was confirmed by measuring several samples).
One can also see from Fig. 2(c) that the $\varepsilon$-band bottom and the $\eta$-band top at M are shifted downward (by $\sim$20 meV) upon Nb substitution.
These results suggest that the Nb substitution affects both the Sb- and V-derived bands.
In contrast to the downward shift of the $\alpha,\varepsilon$, and $\eta$ bands, the other bands near $E_{\textrm{F}}$ (e.g. $\gamma$ and $\delta$) look relatively stationary.
Since this necessitates a more accurate measurement, we have performed ARPES experiments with higher resolution ($\Delta E = 7$ meV) with the He-I$\alpha$ line ($h\nu = 21.218$ eV).
Plots of ARPES intensity along the MK cut for $x = 0, 0.03,$ and 0.07 are shown in Figs. 2(f)-2(h), respectively.
One can commonly see the $\delta$ band which crosses $E_{\textrm{F}}$ at the midway between M and K, while the intensity distribution is different; the intensity maximum is located around the M point for $x = 0$ while it is slightly away from the M point for $x = 0.03$ and 0.07.
This suggests a finite difference in the Fermi wave vectors ($k_{F}$\textquoteright s).
To see the difference in the $\delta$-band dispersion, we have performed numerical fittings to the EDCs at each $x$ (a representative fitting result is shown in the inset).
A direct comparison of the extracted band dispersion in Fig. 2(i) signifies that the experimental band dispersion at $E_{\textrm{B}} > 40$ meV is rather insensitive to the change in $x$, whereas that within 40 meV of $E_{\textrm{F}}$ exhibits a systematic variation.
The $k_{F}$ point systematically moves away from M, accompanied by a change in the band slope with increasing $x$.
This indicates a shrinkage of the triangular electron pocket at K with Nb substitution, as opposed to an expansion of the electron pocket at $\Gamma$.
Such an opposite band shift is expected from the isovalent substitution with V and Nb ions that causes no effective carrier doping.

\begin{figure*}[htbp]
\includegraphics[width=165mm]{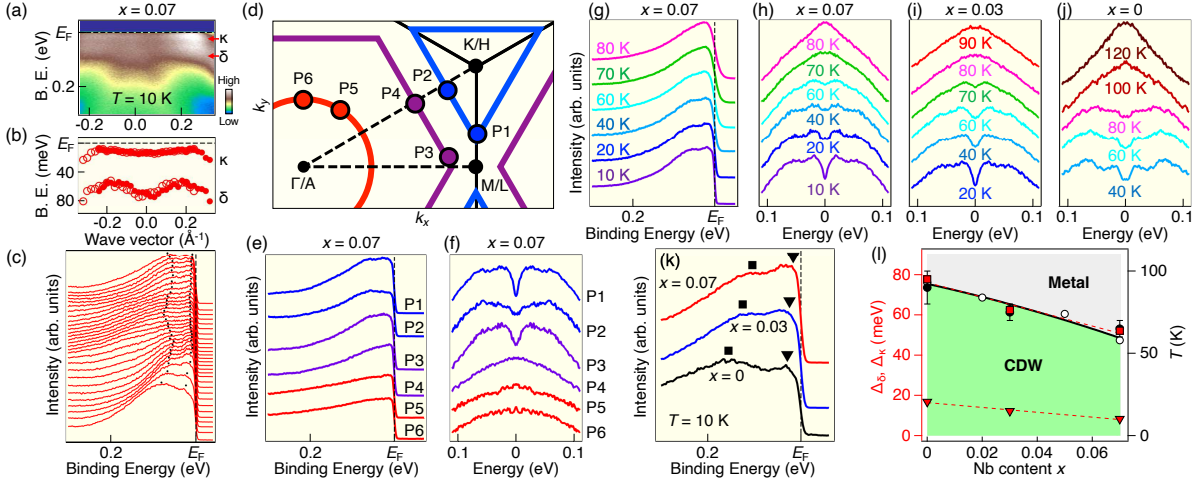}
 \caption{\label{Fig3}
(a) Plot of ARPES intensity near $E_{\textrm{F}}$ along the MK cut for $x = 0.07$ measured at $T$ = 10 K. Red arrows indicate the $\kappa$ and $\delta$ bands.
(b) Experimental band dispersion obtained by numerical fittings to the EDCs in (c) with double Lorentzian peaks multiplied by the FD function convoluted with the resolution function.
(c) Corresponding EDCs of (a).
Dots show experimental dispersion of the $\kappa$ and $\delta$ bands [same as red circles in (b)].
(d) Schematic Fermi surface in the surface BZ together with the $k_{F} $ points (P1-P6) where the EDCs in (e) were obtained.
(e), (f) EDCs and symmetrized EDCs at $T$ = 10 K at P1-P6 in (d).
(g) Temperature dependence of the EDC at P1 for $x = 0.07$.
(h-j) Temperature dependence of the symmetrized EDC at P1 for $x = 0.07, 0.03,$ and 0, respectively.
(k) EDC at P1 at $T$ = 10 K for $x = 0.07, 0.03,$ and 0.
Solid squares and triangles show the CDW gap of the SP band $\Delta_{\delta}$ and the shallow electron pocket $\Delta_{\kappa}$ obtained by the same fittings as that in (b).
(l) Plots of $\Delta_{\delta}$ (red squares) and $\Delta_{\kappa}$ (red triangles) against $x$, compared with the experimental $T_{\textrm{CDW}}$ from ARPES (black circles) and transport (white circles) measurements.}
\end{figure*}

To clarify the influence of Nb substitution on the CDW gap, we have performed high-resolution ARPES measurements below $T_{\textrm{CDW}}$ (= 58 K \cite{NbCVSZhiwei}) for $x = 0.07$ where the CDW is most strongly suppressed.
Figure 3(a) shows the ARPES intensity along the MK cut measured at $T = 10$ K.
Unlike the case above $T_{\textrm{CDW}}$ [Fig. 2(h)], there exist two intensity maxima at $\sim$60 and $\sim$20 meV, associated with the SP band $(\delta)$ with a large CDW gap and a shallow electron band $(\kappa)$ at $k_{z}=\pi$ with a small CDW gap, respectively \cite{NakayamaPRB2021,NakayamaPRX2022} (note that the He-I$\alpha$ photons simultaneously observe the electronic states at $k_{z} = 0$ and $\pi$ planes due to the $k_{z}$ broadening effect).
Both bands show a M-shaped dispersion as better visualized in the experimental band dispersion in Fig. 3(b) which signify a hump-dip-peak structure originating from two types of CDW gaps.
To clarify the \textbf{k} dependence of the CDW gap, we measured EDCs at various $k_{F}$ points [Fig. 3(d)] covering the triangular (P1 and P2), hexagonal (P3 and P4), and circular (P5 and P6) pockets.
As shown in Fig. 3(e), at P1 which is  the $E_{\textrm{F}}$-crossing point of the $\delta$ band, one can see a hump at $\sim$60 meV associated with the large CDW gap $(\Delta_{\delta})$ and a peak at $\sim$20 meV due to the small CDW gap $(\Delta_{\kappa})$.
A similar hump is also observed at P2, whereas it moves closer to $E_{\textrm{F}}$ at P3 and P4 and eventually vanishes at P5 and P6, revealing the strongly $\textbf{k}$-dependent CDW gap \cite{NakayamaPRB2021,LuoNCOM2022,WangarXiv2021}.
This anisotropic gap is clearly visualized in the symmetrized EDCs in Fig. 3(f), revealing a suppression of spectral density of states (DOS) around $E_{\textrm{F}}$ at P1-P3 as opposed to a pile up of DOS at $E_{\textrm{F}}$ at P4-P6.
We have performed temperature-dependent measurements at P1 where the CDW gap takes a maximum.
The EDC and corresponding symmetrized EDC shown in Figs. 3(g) and 3(h), respectively, signify that the suppression of DOS almost recovers at $T$ = 60-70 K close to $T_{\textrm{CDW}}$.
The broad hump also seems to disappear at $T\sim70$ K.
Since this temperature is lower than $T_{\textrm{CDW}}$ of $x = 0$ (93 K), the gap-closing temperature is reduced by the Nb substitution.
This conclusion is corroborated with the temperature dependence of EDCs for $x = 0.03$ ($T_{\textrm{CDW}}$ = 77 K) which signifies the gap closure already at $T$ = 80 K [Fig. 3(i)] in contrast to the EDC for $x = 0$ [Fig. 3(j)] where the DOS suppression is still observed at $T$ = 80 K.

Figure 3(k) shows a comparison of EDCs at $T$ = 10 K measured at P1.
The hump gradually moves toward $E_{\textrm{F}}$ with increasing $x$, reflecting a systematic reduction of $\Delta_{\delta}$.
We have estimated the $\Delta_{\delta}$ and $\Delta_{\kappa}$ values for each $x$ by numerical fittings, and the result together with $T_{\textrm{CDW}}$ are shown in Fig. 3(l). One can identify a general trend that both $\Delta_{\delta}$ and $\Delta_{\kappa}$ are monotonically reduced with increasing $x$, in accordance with the reduction of $T_{\textrm{CDW}}$.
In particular, $x$ dependence of $\Delta_{\delta}$ is well on that of $T_{\textrm{CDW}}$, suggesting the scaling of $T_{\textrm{CDW}}$ and the CDW-gap magnitude.

The present result has an important implication to the mechanism of CDW.
We found in Fig. 3(f) that the maximum CDW gap at $x = 0.07$ opens on the SP band as in the case of $x = 0$ \cite{NakayamaPRB2021,LuoNCOM2022,WangarXiv2021}.
This indicates that the CDW mechanism is unchanged even after the Nb substitution, as naturally expected from the monotonic change in $T_{\textrm{CDW}}$ [Fig. 3(l)].
Moreover, the experimental fact that the reduction of $T_{\textrm{CDW}}$ with Nb substitution is linked to the gradual deviation of the SP from $E_{\textrm{F}}$ supports that the inter-SP scattering stabilizes the in-plane $2\times2$ CDW \cite{NakayamaPRB2021,TanPRL2021,ZhouPRB2021,JiangNM2021}.

Taking into account the essential role of SP to the CDW, the mechanism of intriguing $T_{c}$ enhancement in CVNS is understood as following.
Since the CDW is observed even in the superconducting state, the superconductivity would occur in the metallic $\textbf{k}$ region where the CDW gap is absent \cite{NakayamaPRB2021}.
It is thus suggested that the suppression of CDW with Nb substitution increases the V-derived DOS around $E_{\textrm{F}}$ and contributes to the $T_{c}$ enhancement.
Moreover, since the Sb band maintains the gapless feature even in the CDW phase [Fig. 3(f)], it would also play an important role to the superconductivity.
In fact, observed expansion of the $\alpha$ pocket upon Nb substitution (expanded by 16 \% from $x = 0$ to 0.07) would lead to the increase in DOS at $E_{\textrm{F}}$ and provide more electrons to the superconducting pairing.
Thus, the observed opposite band shift gives rise to better conditions for both the Sb- and V-derived electrons, leading to the cooperative promotion of the superconductivity with higher $T_{c}$.

Finally, it is remarked that the observed opposite band shift in Nb-substituted CVS is different from the band shift reported recently in Ti-substituted CVS \cite{LiuarXiv2021}.
The Ti substitution acts as a hole doping to the crystal, moving the SP away from $E_{\textrm{F}}$ as in the case of Nb-substituted CVS.
Thus, the mechanism of CDW suppression is understood with the same framework for both Nb and Ti substitutions.
On the other hand, the $\alpha$ pocket shrinks with Ti substitution in contrast to the case of Nb substitution.
This shrinkage produces lower DOS at $E_{\textrm{F}}$ and would be responsible for the $T_{c}$ reduction at higher Ti concentrations \cite{LiuarXiv2021,YangarXiv2021}, distinct from the Nb case where the $T_{c}$ monotonically increases with Nb substitution.
Such a critical difference in the behavior of $T_{c}$ is well understood in terms of the difference in the electron filling of the $\alpha$ pocket. 

In conclusion, we reported an ARPES study of $\textrm{Cs(V}_{1-x}\textrm{Nb}_{x})_{3}\textrm{Sb}_{5}$ as a function of $x$.
We uncovered an intriguing change in the band structure upon Nb substitution characterized by the downward shift of the Sb-derived electron band in contrast to the upward shift of the V-derived SP band.
We also found that $T_{\textrm{CDW}}$ becomes higher when the SP band is located at closer to $E_{\textrm{F}}$, pointing to a close connection between the SP and CDW.
We have concluded that the enhancement of $T_{c}$ with Nb substitution is caused by the cooperative expansion of Sb-derived electron pocket and recovery of V-derived density of states at $E_{\textrm{F}}$.

%Acknowledgment
This work was supported by JST-CREST (No. JPMJCR18T1), JST-PRESTO (No. JPMJPR18L7), Grant-in-Aid for Scientific Research (JSPS KAKENHI Grant Numbers JP21H04435 and JP20H01847), KEK-PF (Proposal number: 2021S2-001), and UVSOR (Proposal number: 21-658 and 21-679).
The work at Beijing was supported by the National Key R\&D Program of China Grant No. 2020YFA0308800), the Natural Science Foundation of China (Grants No. 92065109), the Beijing Natural Science Foundation (Grant No. Z210006), and the Beijing Institute of Technology (BIT) Research Fund Program for Young Scholars (Grant No. 3180012222011).
T.K. acknowledges support from GP-Spin.
Z.W. thanks the Analysis \& Testing Center at BIT for assistance in facility support.

%\bibliography{NbCVS}% Produces the bibliography via BibTeX.

%

\end{document}